\documentclass[useAMS,usenatbib]{mn2e}

\title[IC e$^\pm$ pair cascade model for LSI +61$^{\rm o}$ 303]
{Inverse Compton e$^\pm$ pair cascade model for the gamma-ray
production in massive binary LSI +61$^{\rm o}$ 303}
\author[W. Bednarek]{W. Bednarek\thanks{E-mail:
bednar@fizwe4.fic.uni.lodz.pl}\\
Department of Experimental Physics, University of \L \'od\'z,
ul. Pomorska 149/153, 90-236 \L \'od\'z, Poland}
\begin{document}

\date{Accepted . Received ; in original form }

\pagerange{\pageref{firstpage}--\pageref{lastpage}} \pubyear{2006}

\maketitle

\label{firstpage}

\begin{abstract}
We apply an inverse Compton $e^\pm$ pair cascade model for $\gamma$-ray production in the massive 
binary system LSI +61$^{\rm o}$ 303 assuming that electrons are accelerated already inside 
the inner part of the jet launched by the compact object. $\gamma$-ray spectra, affected by 
the cascade 
process, and lower energy spectra, from the synchrotron cooling of the highest energy electrons 
in the jet, are calculated as a function of the phase of this binary system. 
$\gamma$-ray spectra expected in such model have different shape than those ones produced
by electrons in the jet directly to observer.
Moreover, the model predicts 
clear anti-correlation between $\gamma$-ray fluxes in the GeV (1-10 GeV) and TeV ($>200$ GeV) 
energy ranges with the peak of the TeV emission at the phase $\sim$0.5  
(the peak half width ranges between the phases $\sim$0.4-0.9 for the inclination 
of the binary system equal to $60^{\rm o}$, and $\sim$0.4-0.1 for $30^{\rm o}$). 
The fine features of TeV $\gamma$-ray emission (fluxes and spectral shapes) as 
a function of the phase of the binary system are consistent with recent observations  
reported by the MAGIC collaboration. Future simultaneous 
observations in the GeV energies (by the GLAST and AGILE telescopes) and in the TeV energies 
(by the MAGIC and VERITAS telescopes) should test other predictions of the considered model
supporting or disproving the hypothesis of acceleration of electrons already in the inner part of the microquasar jets.  

\end{abstract}

\section{Introduction}

The binary system LSI +61$^{\rm o}$ 303 contains a massive 
Be type star, with a radius  of $r_\star$=13.4 R$_\odot$ and a surface temperature $T_{\rm s} = 2.8\times 10^4$ K, and a compact object on a close  orbit characterised by 
a semi-major axis $a = 5.3 r_\star$, an ellipticity $e = 0.72$, and
an azimuthal angle of the observer in respect to the periastron passage 
$\omega = 70^{\rm o}$ (Casares et al.~2005).
The inclination angle of the binary system is not well constrained by 
the observations and ranges from  $25^{\rm o} < i < 60^{\rm o}$ for the case of a neutron 
star and $\theta< 25^{\rm o}$ for a black hole. In the radio wavelengths, the 
observed asymmetric extended structure is usually interpreted as due to relativistic radio jets, 
with an estimated  speed $\sim$0.6c (Massi et al. 2004). This radio source has been related 
by Gregory \& Taylor~(1978) to the  COS B $\gamma$-ray source CG135+01 (Hermsen et al. 1977). 
CG135+01 has been also detected by the EGRET instrument 
above 100 MeV (3EG J0241+6103, Hartman et al. 1999) with a hard spectrum (with a photon spectral index 
$2.05\pm 0.06$, Kniffen et al. 1997), and by the COMPTEL detector in the energy range $\sim$0.75 
to 30 MeV (with a photon spectral index $1.95^{+0.2}_{-0.3}$, van Dijk et al. 1996). 
The analysis of different EGRET observations
shows evidences of variability (Tavani et al. 1998, confirmed by Wallace 
et al. 2000) presenting a modulation with timescales similar to the orbital period of  
LS I +61$^{\rm o}$ 303 and the emission maxima likely near the periastron passage and phase 0.5
(Massi 2004).
The Whipple group (Hall et al.~2003, Fegan et al.~2005) set an upper limit on the TeV flux from 
this source, $0.88\times 10^{-11}$ cm$^{-2}$ s$^{-1}$ above 500 GeV and 
$1.7\times 10^{-11}$ cm$^{-2}$ s$^{-1}$ above 350 GeV, which is clearly below an extrapolation 
of the EGRET spectrum.
Very recently, the MAGIC telescope observed TeV $\gamma$-ray emission at energies above 
$\sim$200 GeV with the power law spectrum (spectral index -2.6) and extending up to a few TeV 
(Albert et al.~2006). This emission is observed after the periastron passage of the compact object.
$\gamma$-ray emission from LSI +61$^{\rm o}$ 303 has been interpreted in terms of the inverse Compton (IC) scattering model in which stellar radiation is scattered by electrons accelerated  
on the shock wave either created in collisions of the stellar and pulsar winds (e.g. Maraschi \& Treves 1981), 
or by electrons accelerated in the jet and scattering of synchrotron and stellar radiation  
(e.g. Bosch-Ramon \& Paredes~2004). The alternative hadronic model for this source has been also 
discussed by Romero, Christiansen \& Orellana~(2005). 
The variable GeV gamma-ray emission reported from LSI +61$^{\rm o}$ 303
by EGRET can be naturally explained provided that the production region of 
$\gamma$-rays 
lays inside the binary system, i.e. not far away from the massive star. Then, 
the optical depths for accelerated electrons (and produced $\gamma$-rays) are large enough for  
developing the IC $e^\pm$ pair cascade in the radiation field of the massive star 
(Dubus 2006, Bednarek~2006 (B06)). For the geometry of the considered binary system, 
the radiation field is strongly anisotropic if considered with respect to the injection place 
of the electrons. Such general type of anisotropic IC $e^\pm$ pair cascades in the radiation of 
the massive stars have been considered in the past by 
Bednarek~(1997, 2000) and Sierpowska \& Bednarek~(2005), and recently applied to the binary 
systems LS 5039 and LSI  +61$^{\rm o}$ 303 by Bednarek~(B06). 
Aharonian et al.~(2006) also consider anisotropic cascades with the application to LS 5039.
The basic features of such cascades differ significantly from the well-known 
IC $e^\pm$ pair cascades  in the isotropic radiation fields studied in detail
in the past (e.g. Protheroe~1986, Honda~1989, Protheroe \& Stanev~1993, Biller~1995). 

In our previous paper (B06), we considered $\gamma$-ray production in the cascade model 
assuming that electrons are injected isotropically at some distance from the massive star.
Based on those calculations, we were able to predict at which orbital phases
the largest fluxes of TeV $\gamma$-rays should be expected.
In fact, these predictions have been successfully tested by the MAGIC observations 
(Albert et al. 2006). In this paper, based on the anisotropic cascade calculations (B06), 
we discuss a more specific model 
for the $\gamma$-ray production  in LSI +61$^{\rm o}$ 303. Here we assume that electrons are accelerated on shocks waves propagating along the jet launched from the vicinity of the compact object. They are injected along the jet with specific injection efficiency which does not depend on the phase of the binary system. We do not consider the effects of possible changes of the acceleration efficiency related to, e.g. variations in the accretion rate onto the compact object and neglect the possible radiation from the equatorial disk around the massive Be star in respect to its emission from the surface.
In the present code we include the synchrotron losses of electrons in the jet which were not considered in previous work (B06).

\section{IC $e^\pm$ pair cascade model} 

In the simple model considered in this paper it is assumed that electrons are accelerated 
continuously along the jet propagating from the compact object perpendicularly to the plane of 
the binary system. We can apply a simple scaling of the magnetic field along the jet assuming 
its conical structure with the fixed opening angle $\alpha$,
\begin{eqnarray} 
B(x) = B_{\rm d}\eta/(1 +\alpha z)\approx B_{\rm d}\eta /(\alpha z)~~~{\rm 
for~~~z\gg 1/\alpha},
\label{eq1}
\end{eqnarray}
where $\alpha$ is assumed of the order of $0.1$ rad, the scaling factor $\eta < 1$,
and $z$ is the distance along the jet from its base. We assume $B_{\rm d}\approx 3\times 10^5$ G,
i.e. the order of magnitude applied also in other papers (e.g. Bosch-Ramon et al.~2006).
The acceleration rate of electrons (with the Lorentz factors $\gamma$) in the jet can be parametrised by,
\begin{eqnarray}
\dot{P}_{\rm acc}(\gamma) = \xi c E/r_{\rm L}\approx 10^{13}\xi B~~~{\rm eV~s^{-1},} 
\label{eq2}
\end{eqnarray}
\noindent
where $E$ is the energy of accelerated electrons, $\xi$ is the acceleration
efficiency, $r_{\rm L} = E/eB$ is their Larmor radius, 
$B$ is the magnetic field at the acceleration site (in Gauss), 
$e$ is the electron charge, and $c$ is the velocity of light. 
The basic role in the above formula is played by the acceleration efficiency of particles $\xi$.
The theory of shock acceleration process estimates the value of $\xi$ as $\sim 0.1(v/c)^2$,
i.e. $\sim 0.04$ for $v = 0.6c$ (Malkov \& Drury~2001).
On the other hand, the presence of relativistic leptons with energies $\sim 10^{15}$ eV inside the Crab 
Nebula suggests that $\xi$ should be not very far from unity for relativistic winds.
Therefore, we consider values of $\xi$ in the range $0.3-0.03$.

During the acceleration process, electrons lose energy mainly by the synchrotron 
and the inverse Compton (IC) processes. We are interested in the situations in which
electrons can be accelerated to TeV energies. This is not easy to obtain relatively close 
to the base of the jet where the magnetic field is large and the synchrotron losses
very strong. Therefore, in the case of the IC process we consider distances from the accretion disk 
for which the energy losses on the massive star radiation dominate and the IC losses on the disk radiation
and the synchrotron photons can be neglected. 

Let us estimate the maximum energies of electrons in such shock acceleration process
allowed by the energy losses. The energy loss rate by the synchrotron process is given by,
\begin{eqnarray}
{\dot P}_{\rm syn}(\gamma) = {{4}\over{3}}\pi \sigma_{\rm T} c U_{\rm B}\gamma^2 
\approx 2.07\times 10^{-3}B^2\gamma^2~~~{\rm eV~s^{-1},} 
\label{eq3}
\end{eqnarray}
\noindent
where $U_{\rm B} = B^2/8\pi$ is the energy density of the magnetic field, 
$\gamma$ is the Lorentz factor of electrons, and $\sigma_{\rm T}$ is the Thomson cross 
section. Electrons lose energy on the IC process in the Thomson (T) regime, if their
Lorentz factors are, 
\begin{eqnarray}
\gamma\ll \gamma_{\rm T/kN} = m c^2/3k_{\rm B}T\approx 2\times 10^5/T_4,
\label{eq3a}
\end{eqnarray}
(where $T = 10^4T_4$ K is the surface temperature of the massive star which determines the black body spectrum, and $k_{\rm B}$ is the Boltzman constant),
and in the Klein-Nishina (KN) regime, for the Lorentz factors $\gamma\gg \gamma_{\rm T/KN}$.
The energy loss rate in the Thomson  regime is given by,
\begin{eqnarray}
{\dot P}_{\rm IC}^{\rm T}(\gamma) = {{4}\over{3}}\pi \sigma_{\rm T} c 
U_{\rm rad}\gamma^2 \approx 3.8T_4^4\gamma^2/r^2~~~{\rm eV~s^{-1},} 
\label{eq4}
\end{eqnarray}
\noindent
where $U_{\rm rad} = 4.7\times 10^{13}T_4^4/r^2$ eV cm$^{-3}$, and $r$ is the distance
to the centre of the massive star in units of its radius $r_\star$.
The energy loss rate in the KN regime depends only logarithmically on the Lorentz 
factor of electrons. We approximate these losses by (see e.g. Blumenthal \& Gould 1970), 
\begin{eqnarray}
{\dot P}_{\rm IC}^{\rm KN}(\gamma)\approx 
{\dot P}_{\rm IC}^{\rm T}(\gamma_{\rm T/KN})\ln{(4k_{\rm B}T\gamma/mc^2-2)}. 
\label{eq5}
\end{eqnarray}
\noindent 
The maximum energies of accelerated electrons are determined by the balance between
the acceleration mechanism and by the most efficient mechanism for the energy losses.
The comparison of Eqs.~\ref{eq2} and~\ref{eq3} gives the maximum allowed Lorentz factors
of electrons due to the synchrotron energy losses,
\begin{eqnarray}
\gamma_{\rm syn}\approx 6\times 10^7(\xi/B)^{1/2}. 
\label{eq6}
\end{eqnarray}
The maximum energies allowed by IC process in the Thomson regime are (from Eq.~\ref{eq2} and~\ref{eq4}), 
\begin{eqnarray}
\gamma_{\rm IC}^{\rm T}\approx 1.6\times 10^6(\xi B)^{1/2}r/T_4^2, 
\label{eq7}
\end{eqnarray}
and in the KN regime are (from Eq.~\ref{eq2} and~\ref{eq5}),
\begin{eqnarray}
\gamma_{\rm IC}^{\rm KN}\approx 1.5\times 10^5(2+e^{(71\xi B r^2/T_4^2)})/T_4. 
\label{eq8}
\end{eqnarray}
Note that the limit given by Eq.~\ref{eq7} is only valid provided that 
$\gamma_{\rm IC}^{\rm T}<\gamma_{\rm T/KN}$, i.e when the parameters of the jet and the massive star fulfil the following condition, $T_4/r > 8(\xi B)^{0.5}$ (from Eq.~\ref{eq3a} and~\ref{eq7}). In any other case, the limit given by Eq.~\ref{eq8} has to be taken.  

\begin{figure}
\vskip 7.7truecm
\includegraphics{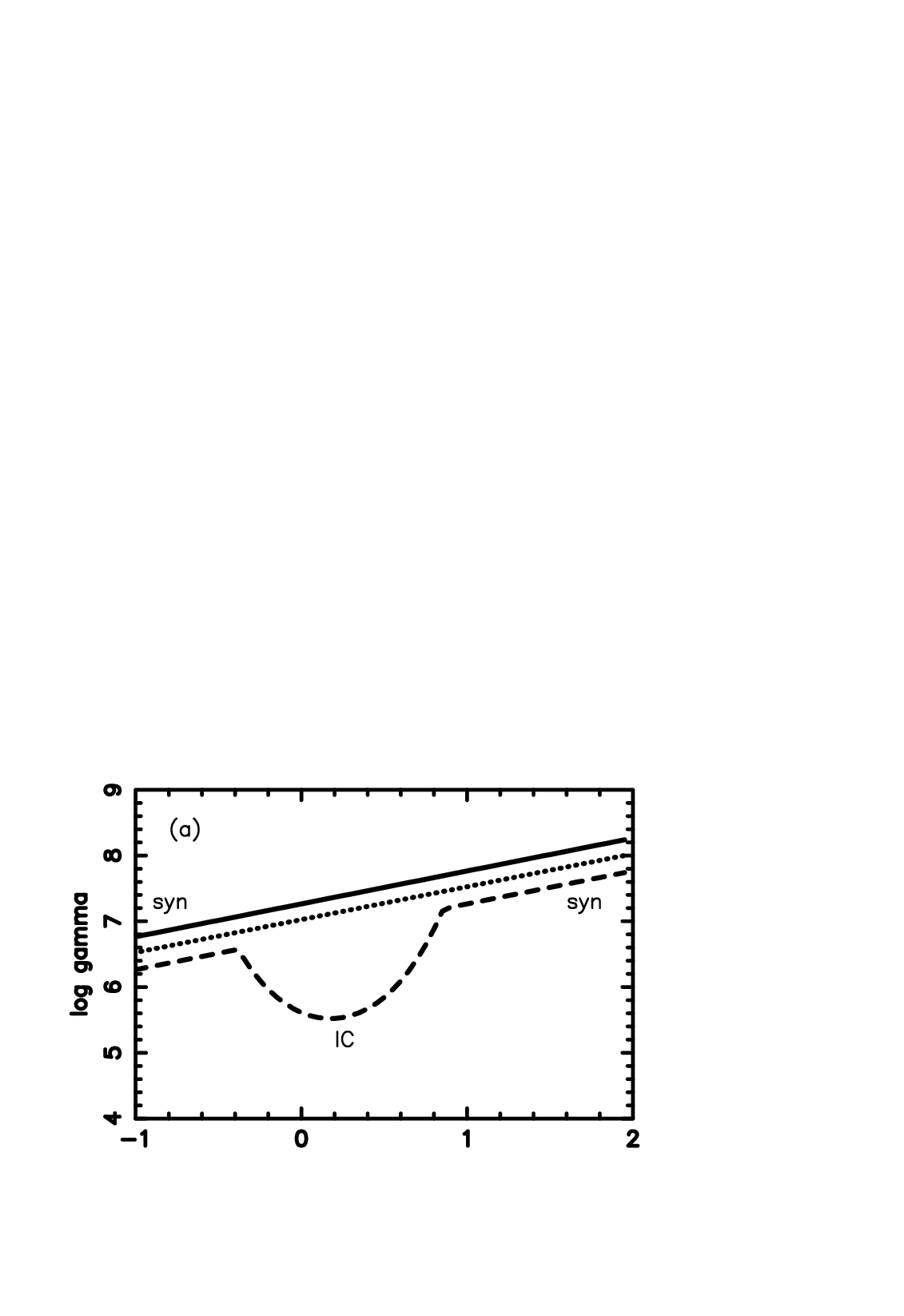}
\includegraphics{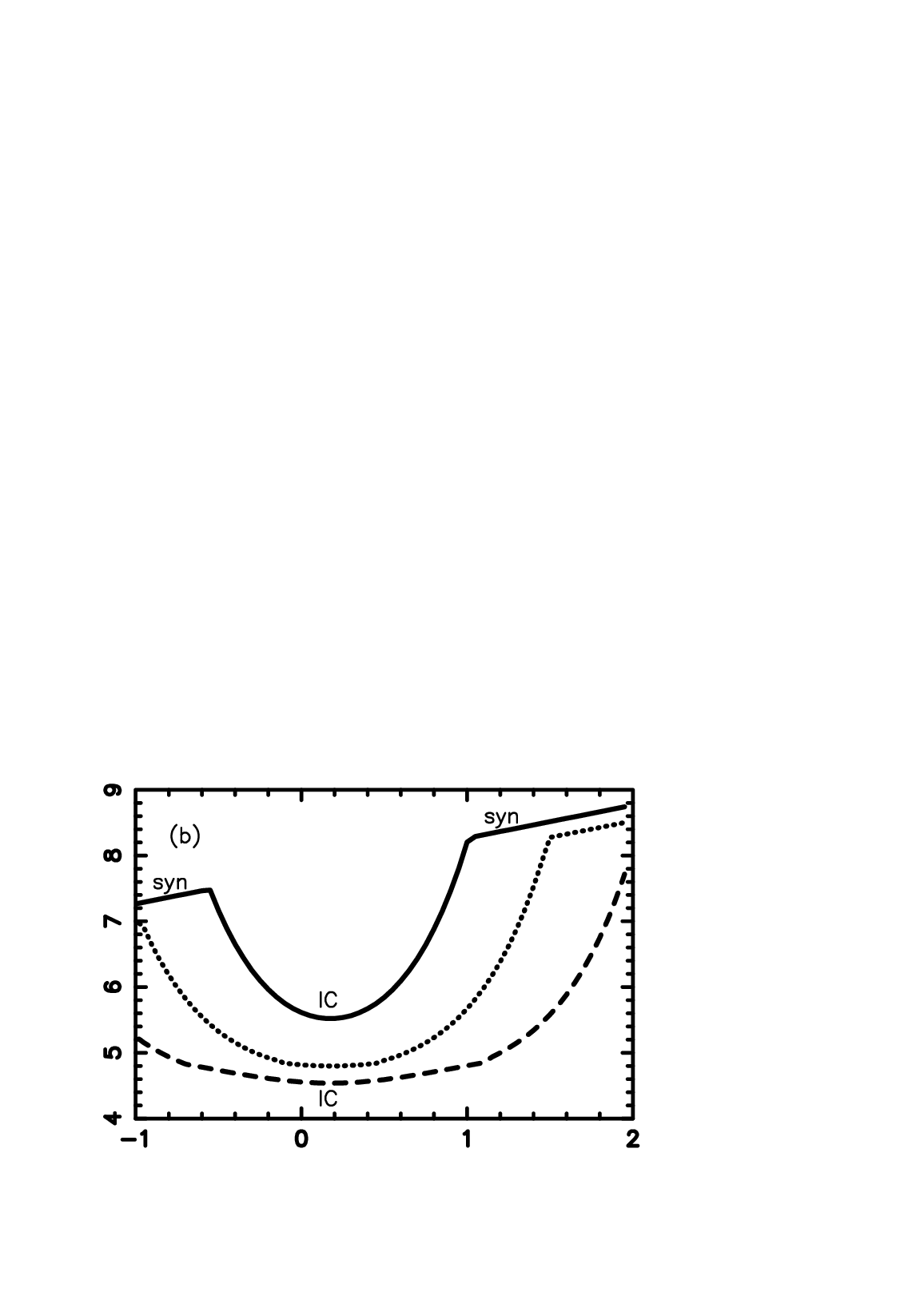}
\includegraphics{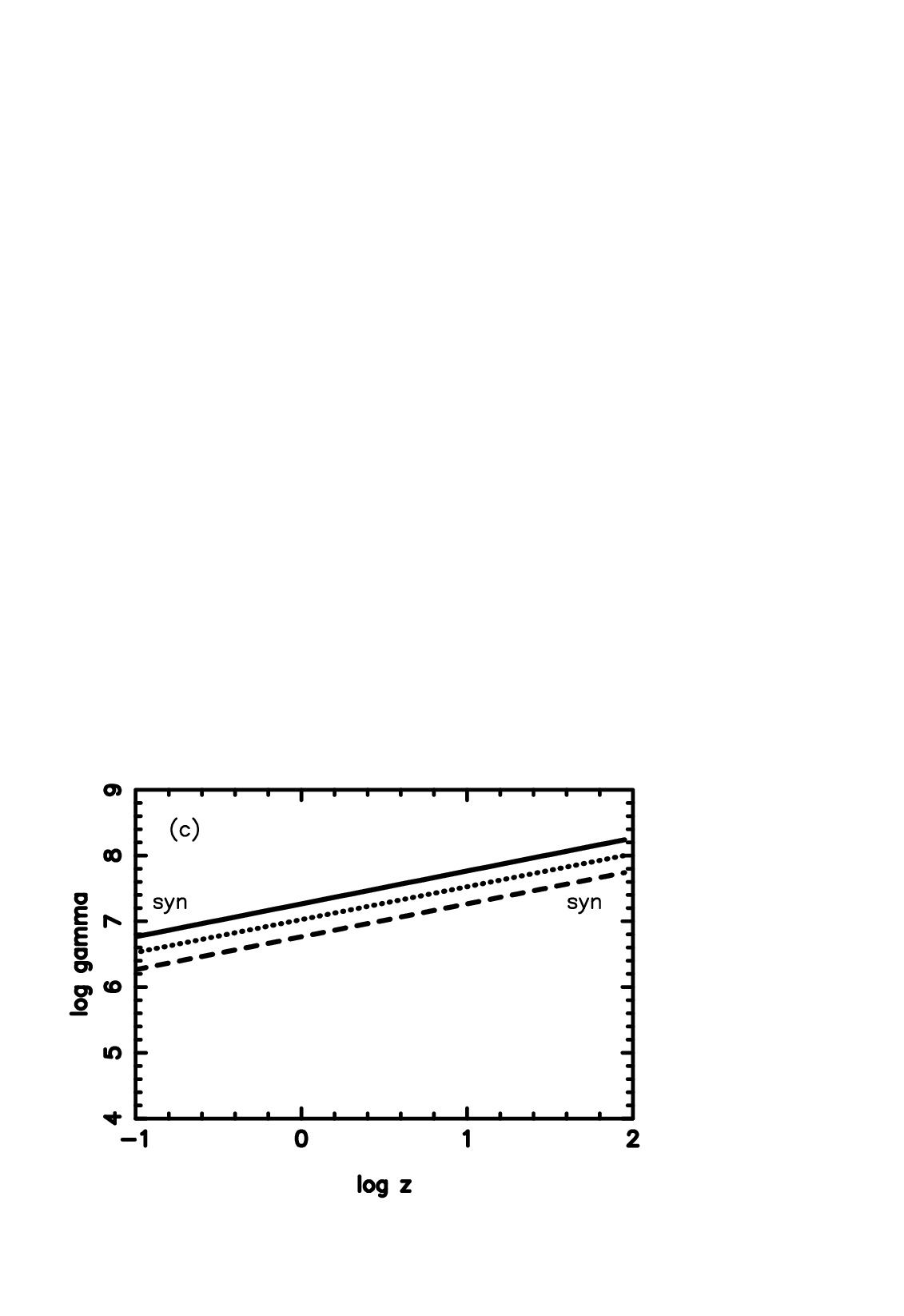}
\includegraphics{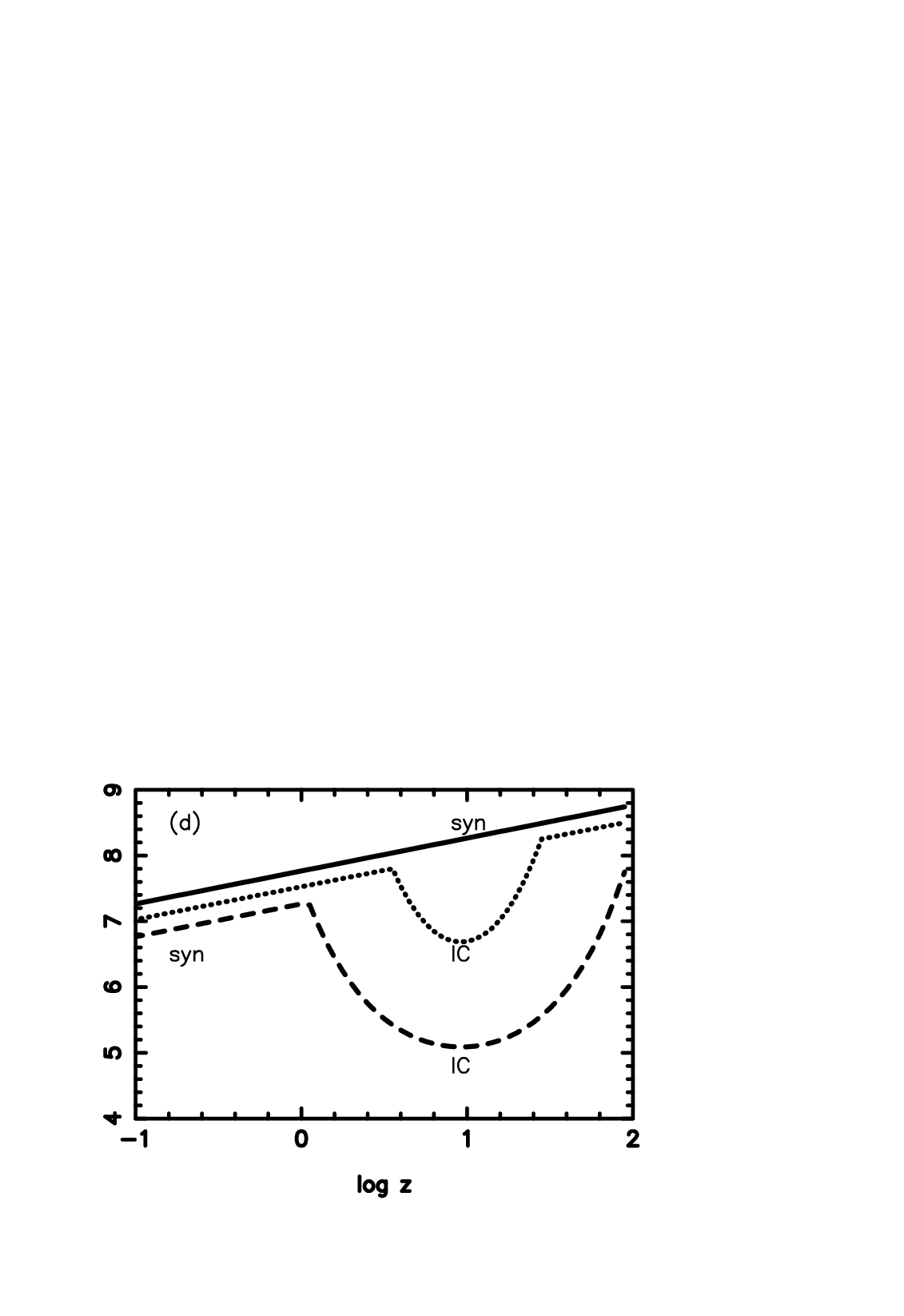}
\caption{The maximum Lorentz factors of electrons as a function of the distance, $z$, 
from the base of the jet (measured in units of the stellar radius), for different 
acceleration efficiencies: $\xi =0.3$ (full curves), $\xi =0.1$ (dotted), and 0.03 (dashed).
Specific figures show the results for 
two ratios of the magnetic field to accretion disk radiation energy densities 
counted at the base of the jet, equal to $\eta=0.1$ (figures a and c), and 0.01 (b and d), 
for the periastron passage of the compact object (a and b) and for the apastron passage 
(c and d). These maximum Lorentz factors are determined by different energy loss
processes.  The comparison of the acceleration rate with the energy loss rate on
the synchrotron process gives the straight parts of curves with dependence $\propto z^{1/2}$
(marked by {\it syn}). 
At some distances acceleration of electrons is saturated by energy losses on the inverse Compton 
scattering with stellar photons in the Thomson and the Klein-Nishina regimes (parts of curves corresponding to broad parabola dips, marked by {\it IC}).}
\label{fig1}
\end{figure}

The maximum Lorentz factors, to which electrons can be accelerated at a specific part of the jet, are determined by the acceleration efficiency and the energy losses through synchrotron and IC processes. These energy loss processes depend on the conditions in the jet (magnetic field strength, acceleration efficiency), on the parameters of the massive star 
(its radius, surface temperature), and on the distance of the acceleration region from
the massive star.  In Fig.~\ref{fig1}, we plot the maximum Lorentz factors of 
electrons as a function of the distance from the base of the jet (given by either Eq.~\ref{eq6}, or Eq.~\ref{eq7}, or Eq.~\ref{eq8}),
for selected values of the acceleration efficiency, $\xi$, and the ratio, $\eta$, 
at the periastron and the apastron passages of the compact object on its orbit around the massive star. It is clear that the maximum energies of accelerated electrons depend 
not only on the distance from the base of the jet but also 
on the phase of the binary system. They increase with the distance from the base of the jet 
as $\propto z^{1/2}$, when the acceleration process is saturated by synchrotron energy losses
(see also Bosch-Ramon et al.~2006).
However, at the regions of the jet where the magnetic field is relatively weak (farther from the base of the jet but still close to the massive star),  the acceleration process is 
saturated by IC energy losses. Then, the maximum Lorentz factors of electrons
are significantly lower  (note the broad parabolic dips in Fig. 1, 
especially evident during the periastron passage of the compact object).

In order to achieve the maximum Lorentz factors, obtained from the above formulae,
the acceleration process has to occur locally in the jet, i.e. the advection time scale of 
relativistic electrons moving with the jet plasma has to be longer 
than the acceleration time to the maximum energies determined by energy losses.
The advection (escape) time scale of electrons can be estimated by,
\begin{eqnarray}
\tau_{\rm adv}\approx z/v_{\rm j}\approx 50z/v_{\rm j}~~~{\rm s}, 
\label{eq9}
\end{eqnarray}
\noindent
where the velocity of the jet $v_{\rm j}$ is in units of the velocity of the light, and 
the distance along the jet $z$ is in units of the stellar radius 
$r_\star$. The advection time scale introduces the upper limit
on the Lorentz factor of the electrons under the condition of their local acceleration,
\begin{eqnarray}
\gamma_{\rm max}^{\rm adv} = {\dot P}_{\rm acc}\tau_{\rm adv}/(mc^2)\approx 
1.8\times 10^9\xi/v_{\rm j}. 
\label{eq10}
\end{eqnarray}
\noindent
For the distances from the base of the jet down to $z = 100$, and acceleration
efficiencies in the range $\xi = 0.03-0.3$, the acceleration 
of electrons occurs locally since $\gamma_{\rm max}^{\rm adv}$ is larger than the maximum Lorentz factors 
given by Eqs.~\ref{eq6},~\ref{eq7},~\ref{eq8} (shown in Figs.~\ref{fig1}).
Regarding particle cooling, it is dominated by radiative processes for the relevant electron energies, providing the particle local cooling, i.e. close to the acceleration place.

\begin{figure*}
\vskip 7.truecm
\includegraphics{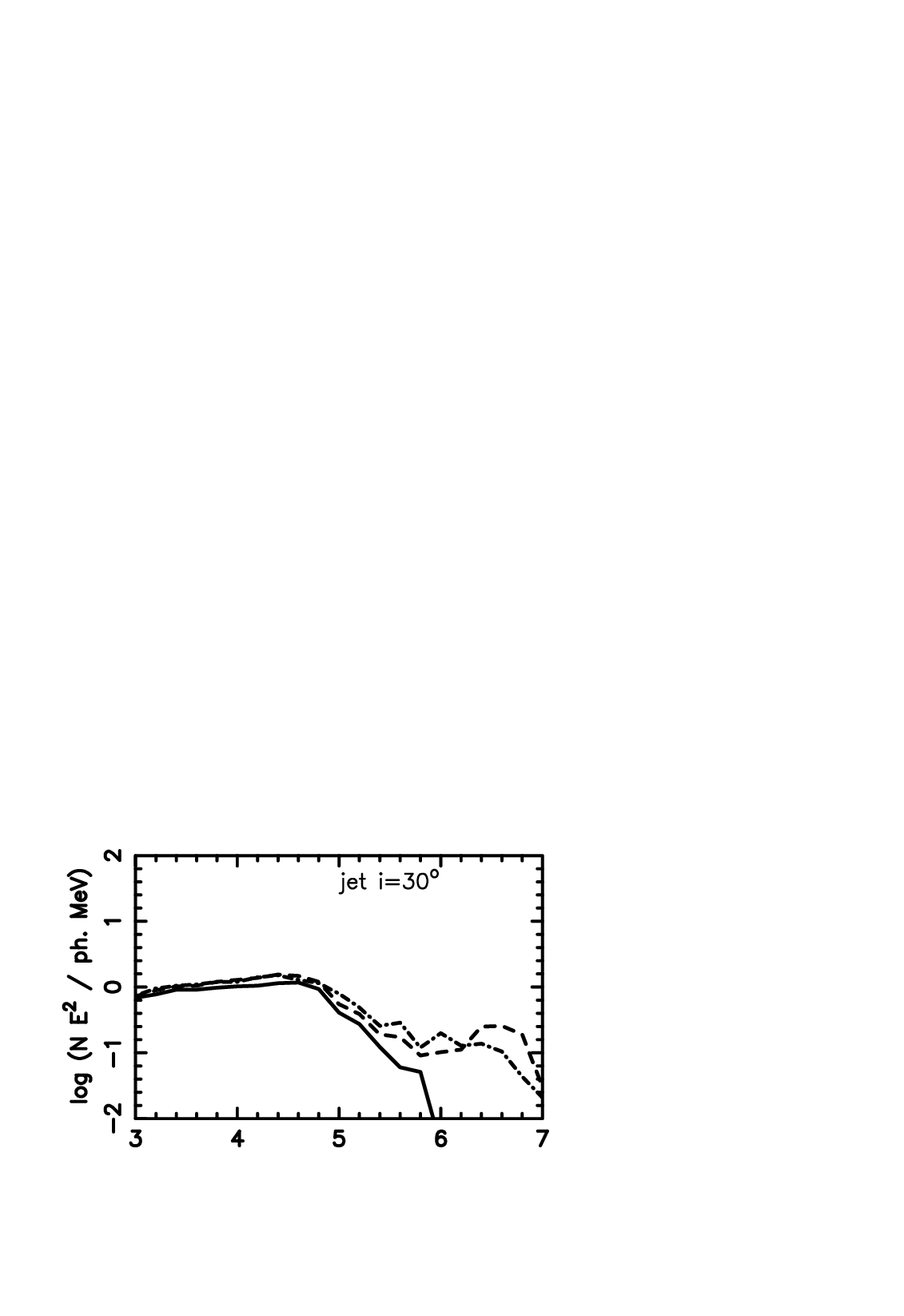}
\includegraphics{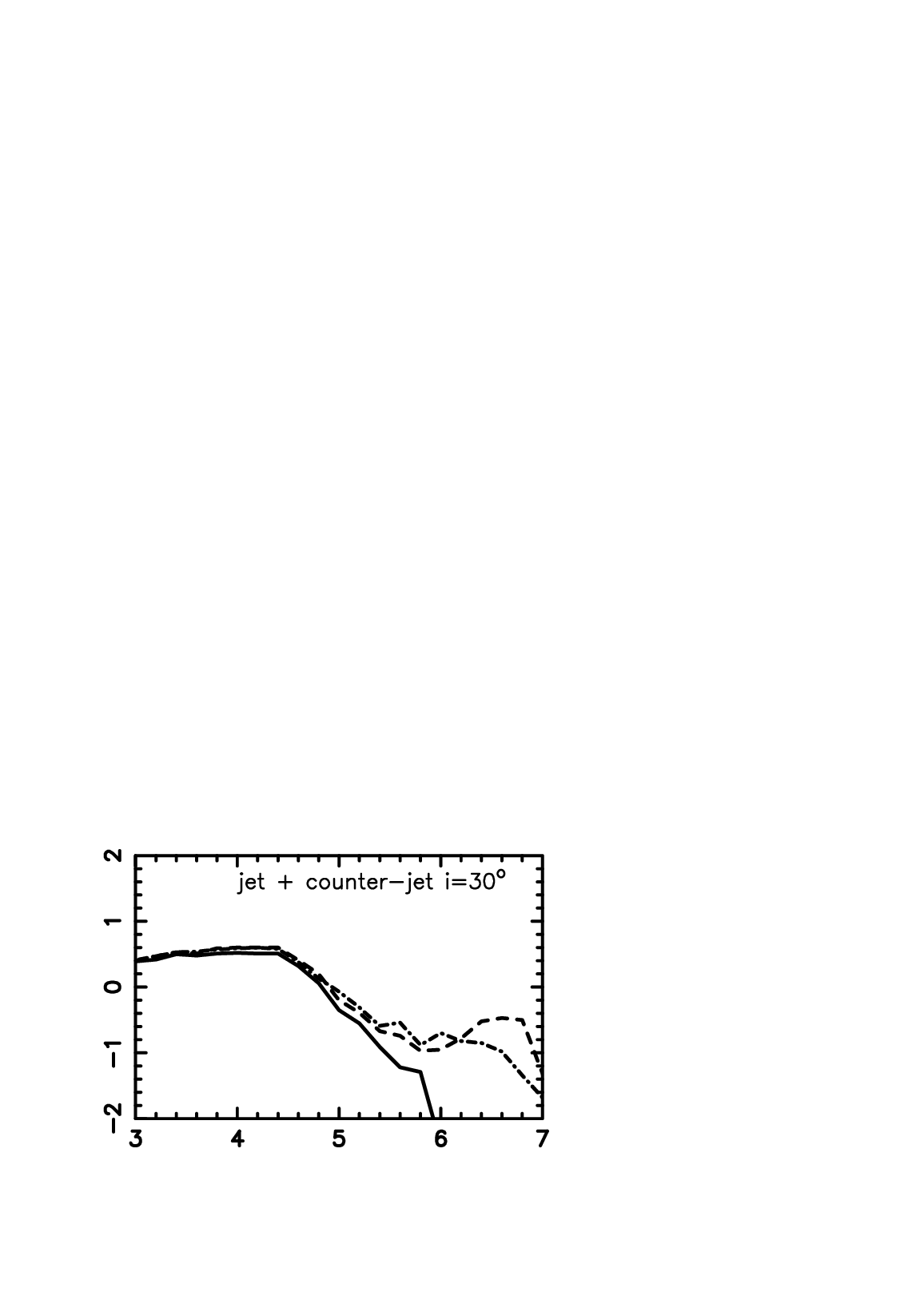}
\includegraphics{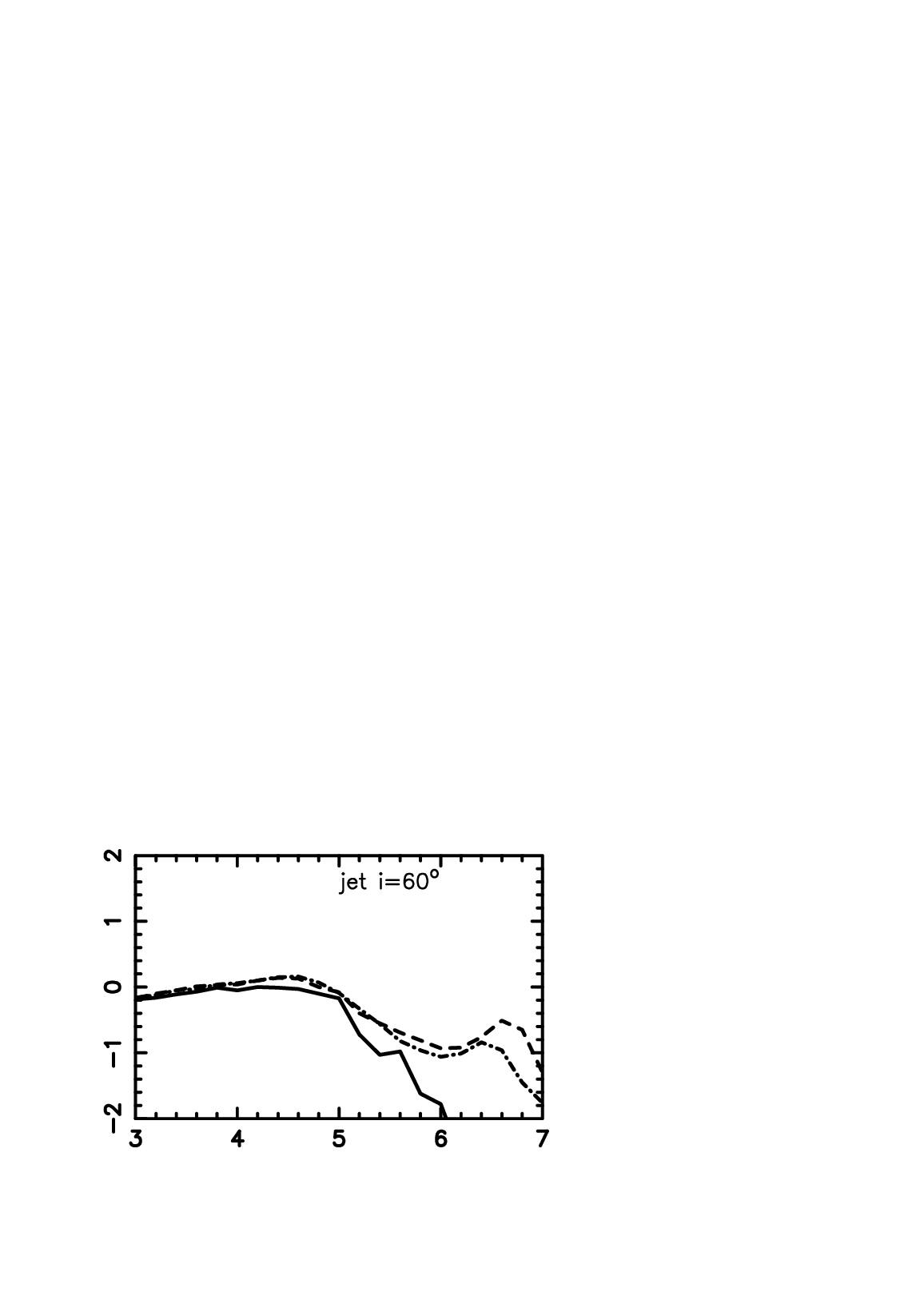}
\includegraphics{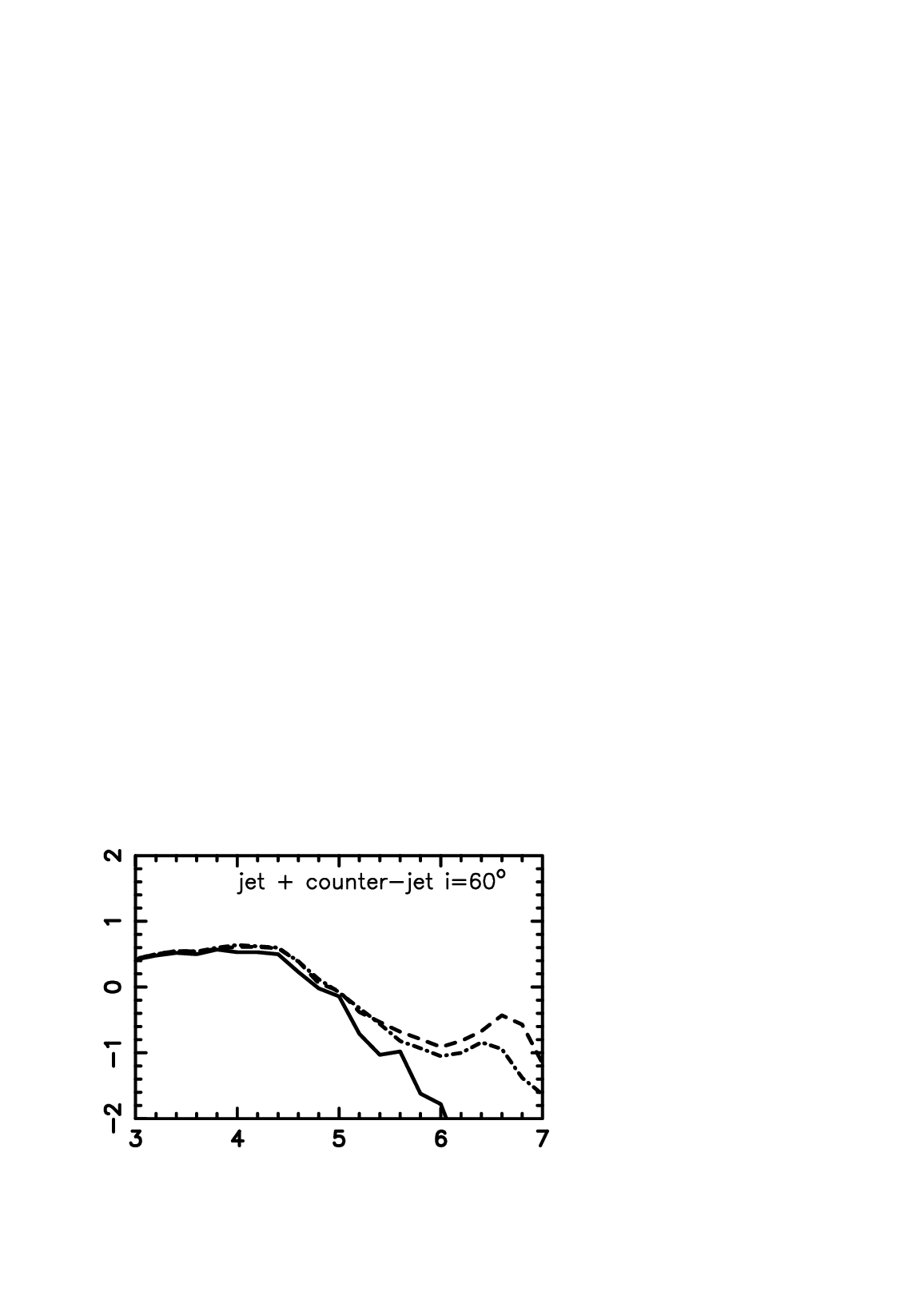}
\includegraphics{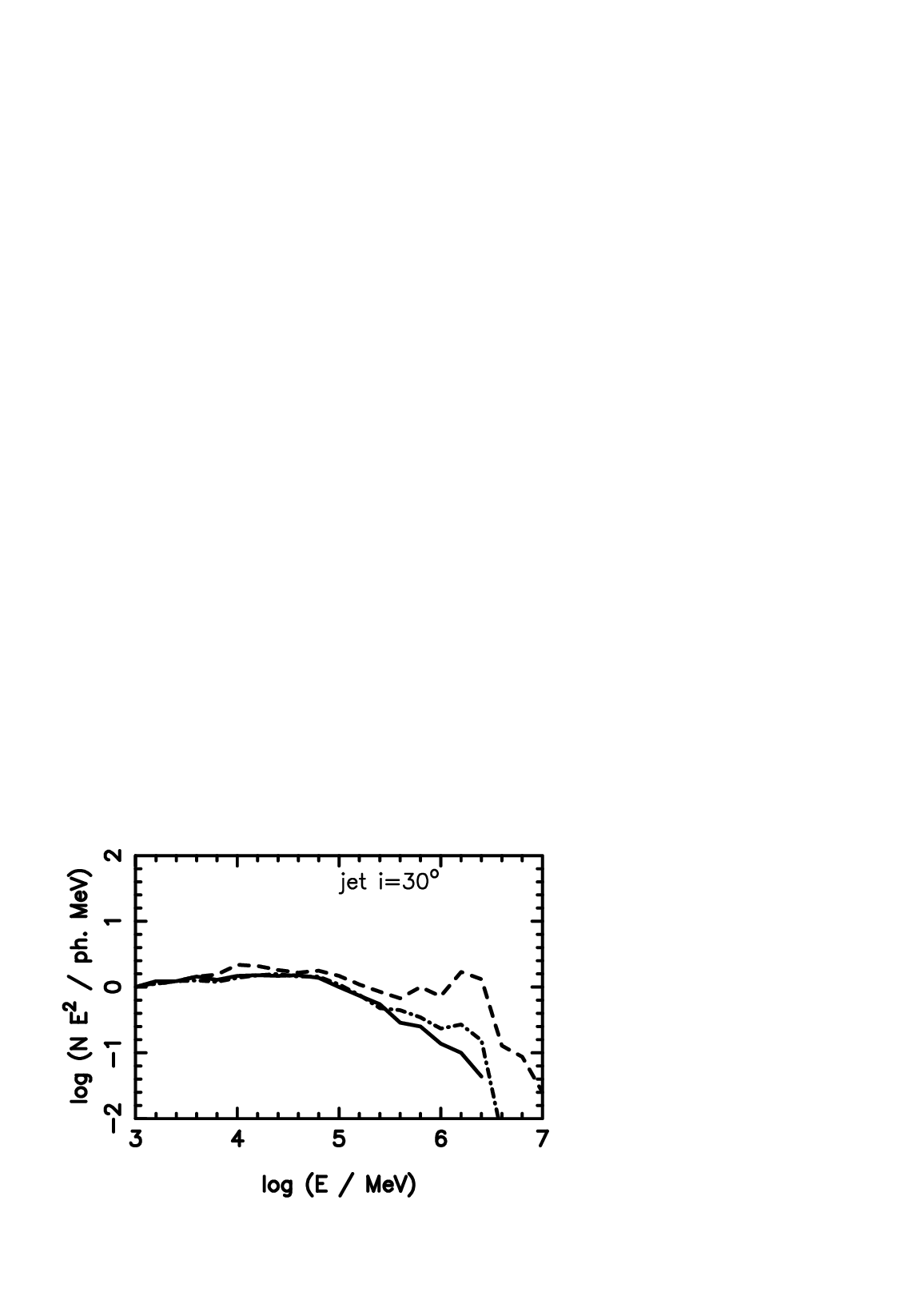}
\includegraphics{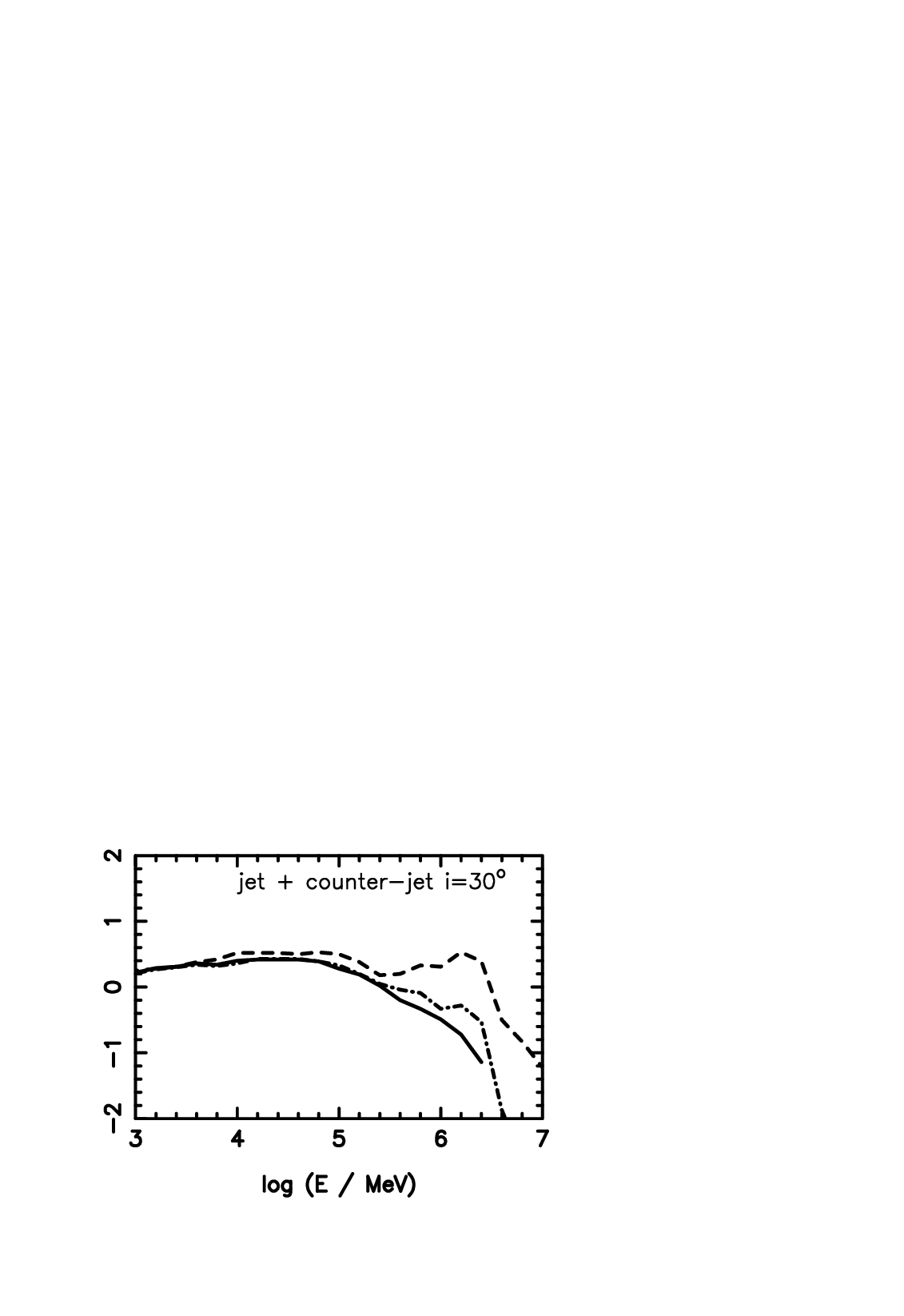}
\includegraphics{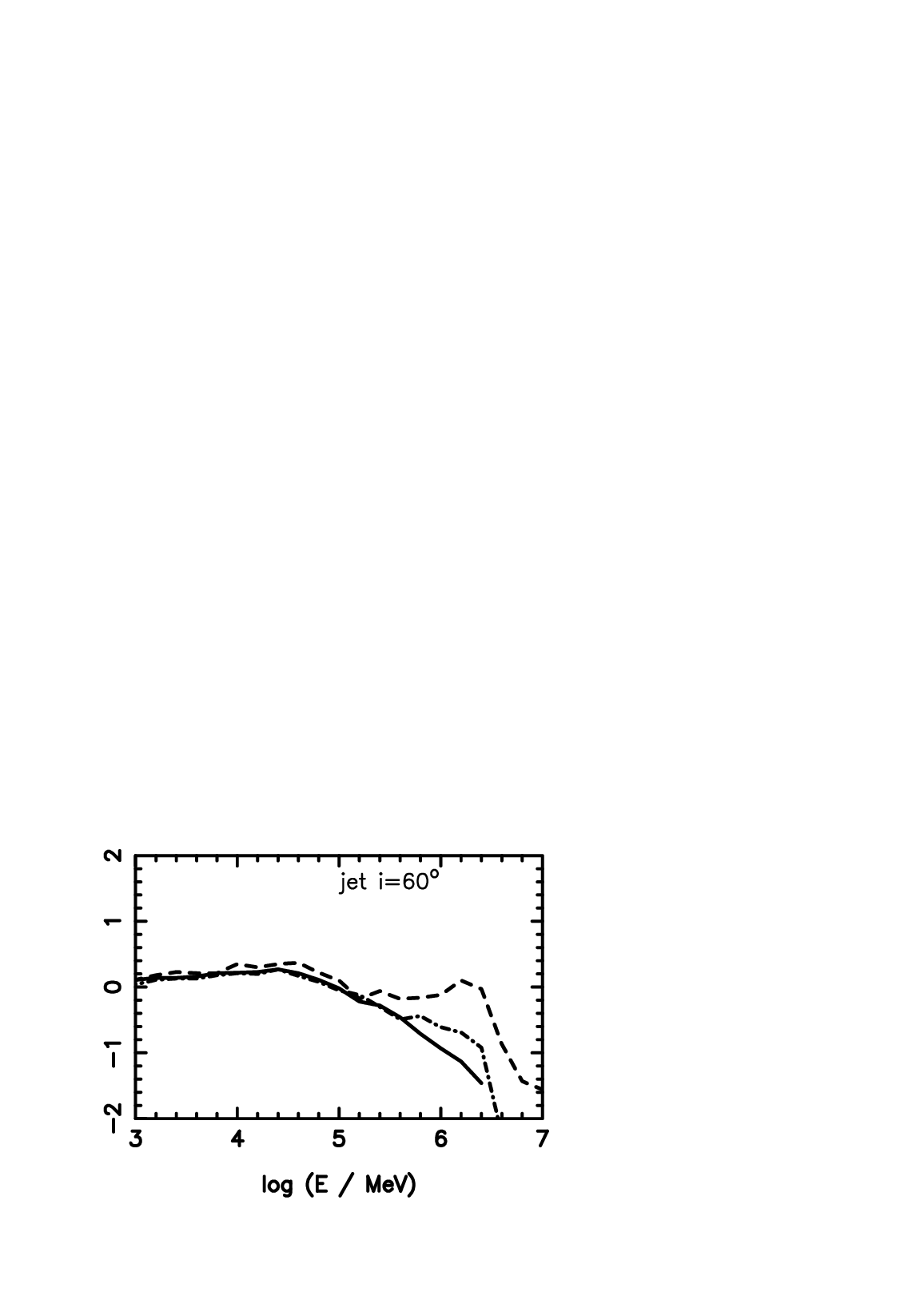}
\includegraphics{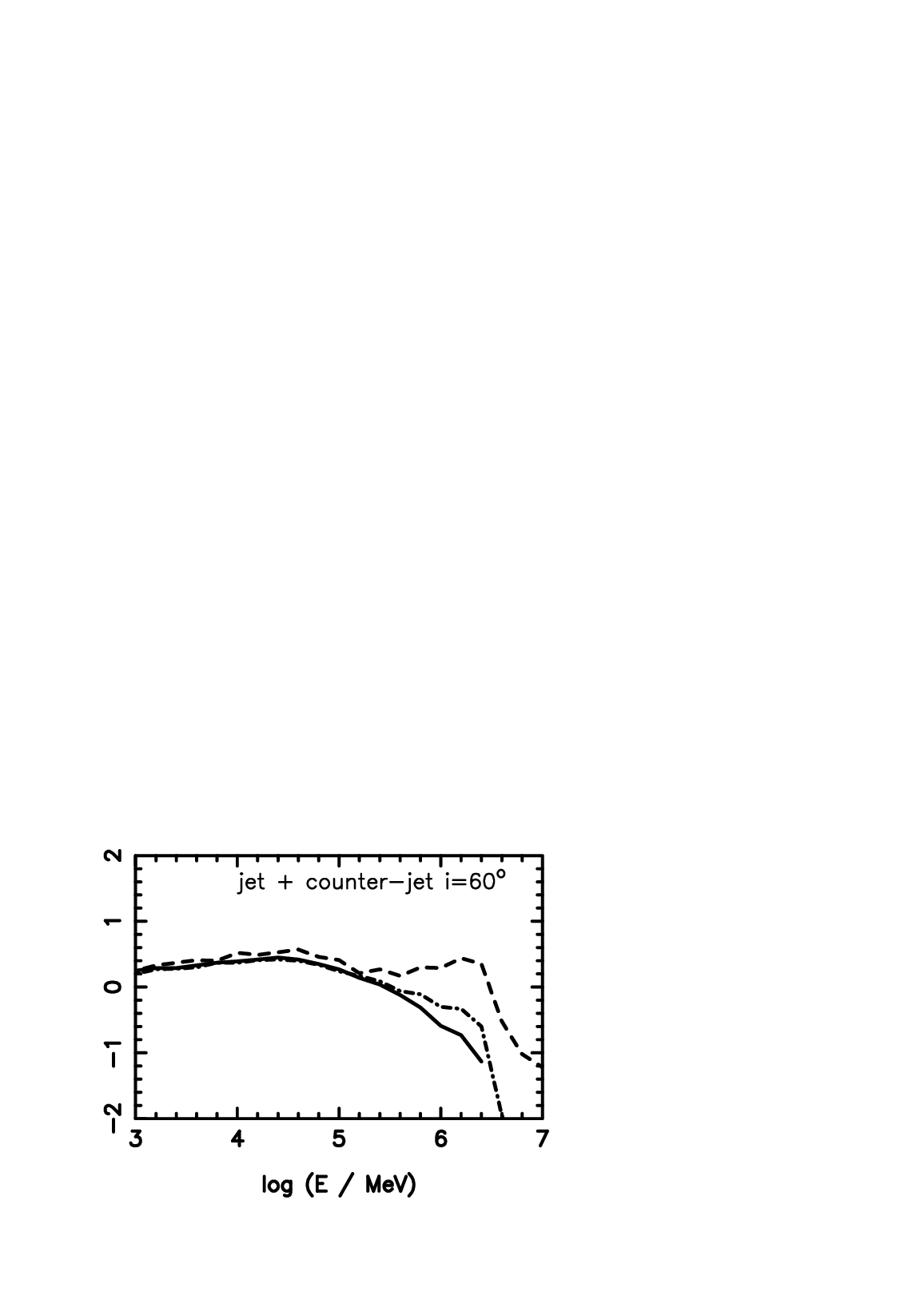}
\caption{Spectral Energy Distributions (SED) 
from cascades initiated by primary electrons injected in the jet with: the efficiency depending on the distance from the base of the jet as $N_{\rm e}(z)\propto 1/z^2$ (independently on 
the phase of the binary system), and differential power law spectrum of electrons with the index equal to -2. The upper figures show the $\gamma$-ray spectra, produced at 
the periastron passage of the compact object for the observer located at the inclination angles
$i = 30^{\rm o}$ and $60^{\rm o}$, from the jet moving towards the hemisphere containing the observer (the {\it jet}) and the sum of the cascade spectra produced by electrons accelerated in the 
{\it jet} and in the {\it counter-jet} (direction opposite the jet). The jets propagate perpendicular to the plane of the binary system. The $\gamma$-ray spectra produced at the apastron
passage of the compact object are shown in the bottom figures.  
The specific spectra are calculated for the acceleration conditions in the jet described by 
the parameters: $\xi =0.03$ and $\eta=0.1$  (full curves),
$\xi =0.3$ and $\eta=0.01$, (dashed), and $\xi =0.3$ and $\eta=0.1$ 
(dot-dashed).} 
\label{fig3}
\end{figure*}

By comparing the rate of energy losses of electrons in the synchrotron process
and the IC process in the Thomson regime (Eqs.~\ref{eq3} and~\ref{eq4}), we find out that at distances from the base of the jet lower than $z_{\rm S/T}\approx 0.57\eta r/T_4^2$, the synchrotron energy losses dominate over the IC energy losses. 
Therefore, essential contribution to the IC spectra comes only
from the part of the jet above $\sim 0.1 r_\star$ since $z_{\rm S/T}$ is typically
lower for the range of considered parameters. 
The part of the jet above $z_{\rm min}$ is considered since losses prevent particles from reaching TeV energies for $z < z_min$. We fix this lower bound on 
$r_{\rm min} = 0.1r_\star$ (for arguments see e.g. B06). From another site, the upper bound on 
this region, $z_{\rm max}$, is introduced by the condition of advection timescales being shorter than radiation timescales.
Based on Eq.~\ref{eq10a}, in all numerical calculations we fix the upper bound of the jet on 
$z_{\rm max} = 10r_\star$. Note moreover, that  electrons might be injected by the acceleration mechanism only at specific range of distances from the base of the jet.
However this possibility can not be reasonably specified at the present state of knowledge.

For the acceleration model defined above, we consider two scenarios for the injection rate of 
relativistic electrons into the jet. The spectrum of injected electrons
is assumed to be of the power law type described by the single spectral index independent 
on the distance from the base of the jet $z$. The injection rate (number of particles at 
specific distance $z$) is assumed to be

\begin{enumerate}

\item constant along the jet from its base at $z_{\rm min}$,

\item drops along the jet as $\propto z^{-2}$.
 
\end{enumerate}

We have shown above that the cooling time scale of electrons which are able to produce $\gamma$-ray photons with energies above 1 GeV is shorter than the advection time
scale of electrons in the jet (see Eq.~(13) and the text below). Since this cooling time scale of electrons  is also much shorter
than the characteristic time scale of the considered picture (determined by the part of the 
orbital period of the binary system), we can assume that all the energy transfered to electrons from the acceleration mechanism at specific time is radiated almost during the same time. 
The spectrum of $\gamma$-rays can be calculated correctly by complete cooling
of the injected electron spectrum.
 
The model with continuous injection rate of electrons produce very flat $\gamma$-ray spectra, 
inconsistent with the upper limits of the Whipple
telescope for LSI +61$^{\rm o}$ 303  (Hall et al. 2003, Fegan~2005)). 
Therefore, we show only the results of calculations for the second injection model of electrons 
into the jet.

Since the optical depths for very high energy $\gamma$-rays in the radiation field
of the massive star in LSI +61$^{\rm o}$ 303 are large,
the $\gamma$-ray spectra, emerging from the binary system toward the observer,
are formed in the IC $e^\pm$ pair cascades occurring in the anisotropic
radiation of the massive stars. We make use of the Monte Carlo cascade code described in detail in previous works 
(Bednarek~2000, B06). However the code has been modified in order to include the effects connected with the synchrotron cooling of electrons injected into the magnetic field of the jet. Therefore, in this work we are also able to show the simultaneous synchrotron photon spectra produced by electrons accelerated in the jet.
When calculating the $\gamma$-ray spectra escaping from the binary system we neglect
the effects of Doppler boosting due to relativistic motion of the jets since for the parameters
typical for LSI +61$^{\rm o}$ 30 (jet speed, $\sim 0.6$  and observation angles $30^{\rm o}$ and $60^{\rm o}$) the Doppler factors are rather small, $D\approx 1.66$ and $1.14$, respectively. 
These relativistic effects change slightly the angular and energy distribution of electrons
in the jet  (as observed from the compact object and the observer rest frame) and so they are neglected in the present calculations.  Note moreover that the possible Doppler
boosting concerns only the primary (first generation) $\gamma$-rays produced by electrons in the jet but not to the secondary cascade $\gamma$-rays.

\section{The cascade $\gamma$-ray spectra}

The distance of the compact object from the massive star in the binary system  LSI +61$^{\rm o}$ 303
changes in the range from $r_{\rm p} = 1.5r_\star$ (periastron) up to
$r_{\rm a} = 9.15r_\star$ (apastron). For the periastron passage the optical depths for $\gamma$-rays injected at most of the hemisphere (on the $e^\pm$ pair production process in $\gamma + \gamma\rightarrow e^\pm$) are larger than unity (see e.g. 
calculations in Dubus 2006, B06).  At the apastron passage the optical depths for the $\gamma$-rays are larger than unity in some range of angles around the massive star (see. e.g.  Bednarek B06). 
Note that isotropic electrons in the jet produce $\gamma$-rays preferencially in directions where they find the largest optical depth.
Therefore,  the absorption of $\gamma$-rays and the interaction of secondary $e^\pm$ pairs have to be taken into account when  calculating the $\gamma$-ray spectra produced
by electrons in the radiation field of the massive star both at the periastron and the apastron passages. We have performed Monte Carlo calculations of the $\gamma$-ray spectra escaping from the binary system at arbitrary
directions, applying the above mentioned models for the acceleration of primary electrons
in the jet, i.e. the injection of electrons along the jet with efficiency dropping from the base of the jet according to $\propto z^{-2}$. 
The IC $e^\pm$ cascade model described in a more detail in Bednarek~(B06) has been applied.
In this model we assume that secondary cascade leptons are isotropized by the random component of the magnetic field in the wind of the massive star. Some motivation for that to possibly occur is discussed in Bednarek~(1997, Sect. 2 of that paper). 
In this cascade process we consider only IC scattering of the radiation coming from the massive 
star neglecting the possible contribution from the soft photon field produced in the accretion disk.
To allow a simple analysis, it is assumed that electrons are accelerated with the power law
spectrum and differential spectral index equal to $-2$. 
The injection of relativistic electrons in the jet as a function of the distance from its base 
has been also made in other papers concerning microquasars (e.g. Bosch-Ramon et al. 2006).
The cut-off in the electron's spectrum depends on the location of the 
acceleration place in the jet and on the phase of the compact object. It is determined by the 
synchrotron, or IC energy losses in the jet and by the acceleration efficiency
(defined by the parameter $\xi$). The values of $\eta$ and $\xi$ are selected in order to allow acceleration
of electrons up to TeV energies (see Fig.~1). Since two jets are expected from a single compact 
object, we  consider the production of $\gamma$-rays in the jet which propagate above the plain of 
the binary system (i.e. the {\it jet} directed towards the observer) and in the jet propagating below the plane of the binary system 
(i.e. the {\it counter-jet}). We note that $\gamma$-rays produced in the counter-jet may have
problems with passing through the matter and radiation of the accretion disk around the compact 
object. The transfer of 
$\gamma$-rays through the region of the disk radiation depends on many unknown accretion disk 
parameters (see e.g. Bednarek~1993) and the column density of matter through the accretion disk in LS I 61$^{\rm o}$+303 is 
unknown. Due to these uncertainties, we show separately $\gamma$-ray spectra from the jet (the case 
when the counter-jet $\gamma$-rays are absorbed) and from the jet + counter-jet (the counter-jet 
$\gamma$-rays are not influenced by the accretion disk).

In order to obtain the first generation of $\gamma$-ray spectra from the injected electron spectrum we cool 
injected electrons up to energies below 3 GeV applying numerical methods. Such approach guaranttee
that all the energy of electrons is transfered to radiation provided that the observation time is much longer than the cooling time of electrons below 3 GeV. Due to this complete cooling of electrons in the jet, the power in electrons accelerated above 3 GeV has be approximately equal to the power in the observed $\gamma$-ray spectrum above 1 GeV. 
The results of the calculations of the $\gamma$-ray spectra from the cascade developed
by these first generation of $\gamma$-rays produced in the jet are shown in Fig.~2 for the case of an observer located at  inclination angles of $30^{\rm o}$ and $60^{\rm o}$, and the compact object at the
periastron (upper figures) and the apastron passages (bottom) (spectrum of electrons normalised to 1 MeV and the injection rate to one electron between $z_{\rm min}$ and $z_{\rm max}$). The $\gamma$-ray spectra, produced by electrons
injected in the jet at the periastron passage, steepen significantly at energies above a few tens of GeV due to the efficient cascading. This cascading effects are smaller when the compact object is close to the apastron. The spectra have similar fluxes in the GeV energy range but differ significantly in the TeV energy range for different parameters $\xi$ and $\eta$. Spectra produced at the periastron have lower fluxes at TeV energies due to the absorption of $\gamma$-rays.
The contribution of the counter-jet to the total spectrum is important only at GeV energies 
at the periastron by increasing the flux calculated from the ${\it jet}$ by a factor of $\sim$3-4.

\begin{figure}
\vskip 12.5truecm
\includegraphics{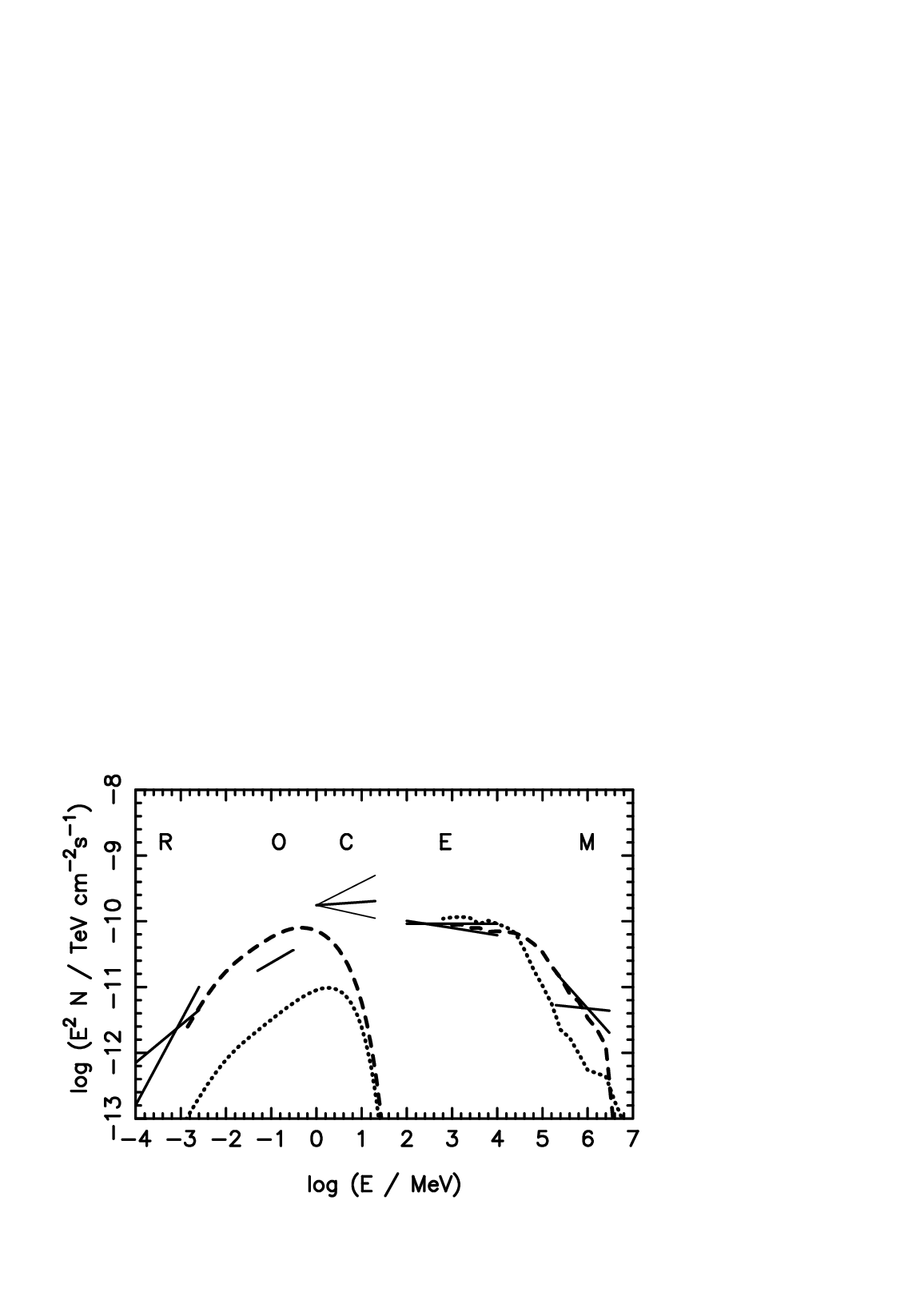}
\includegraphics{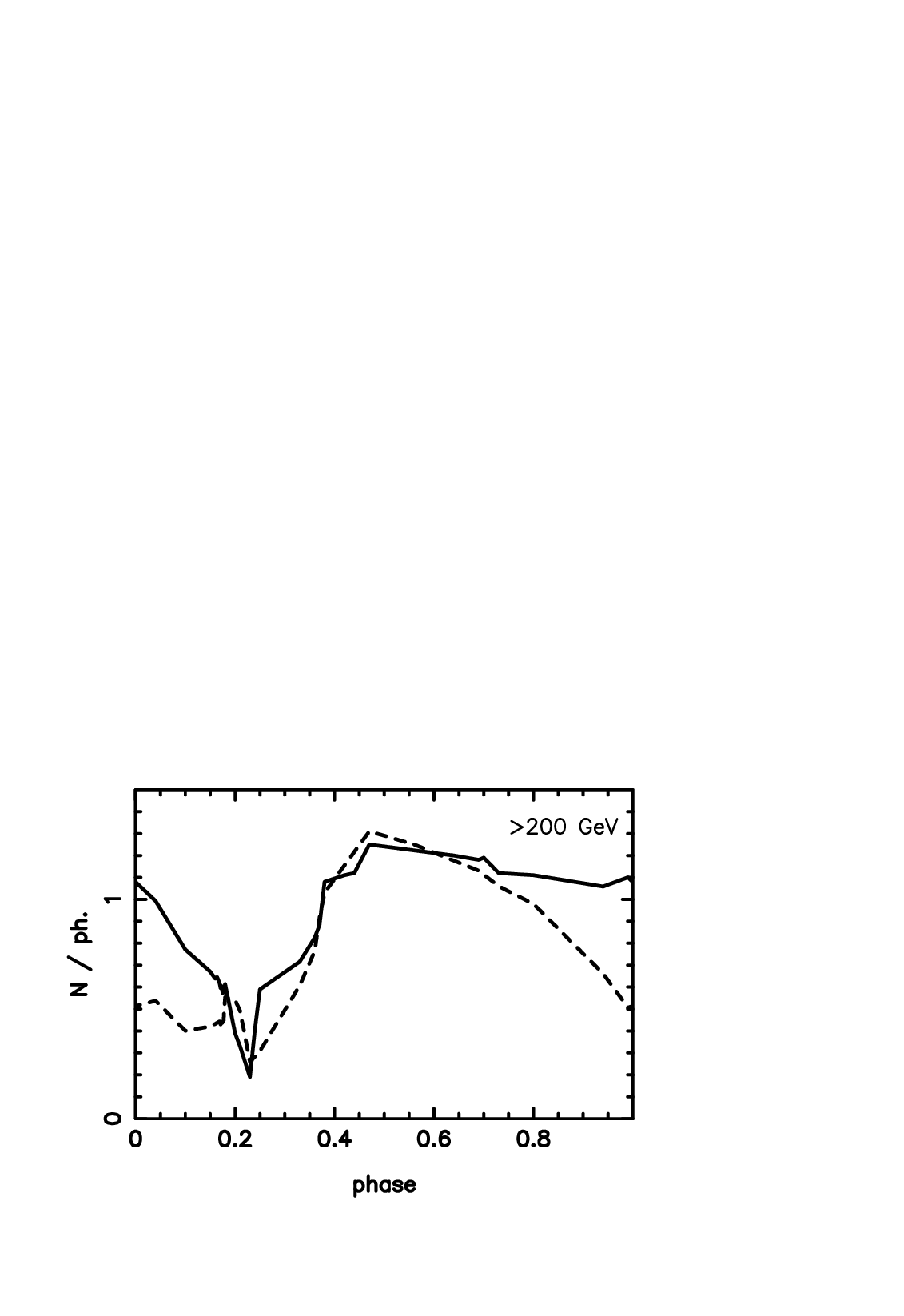}
\includegraphics{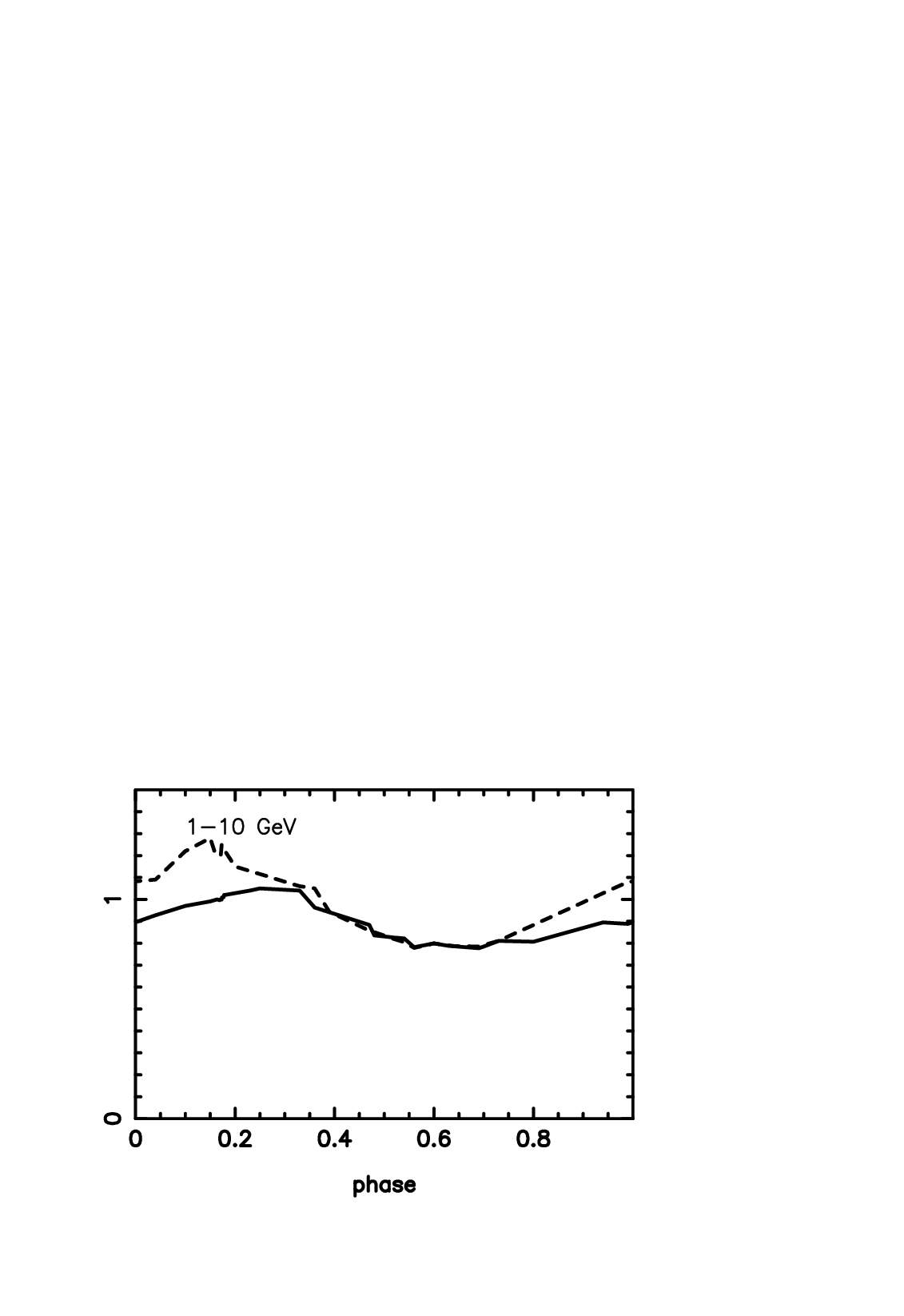}
\includegraphics{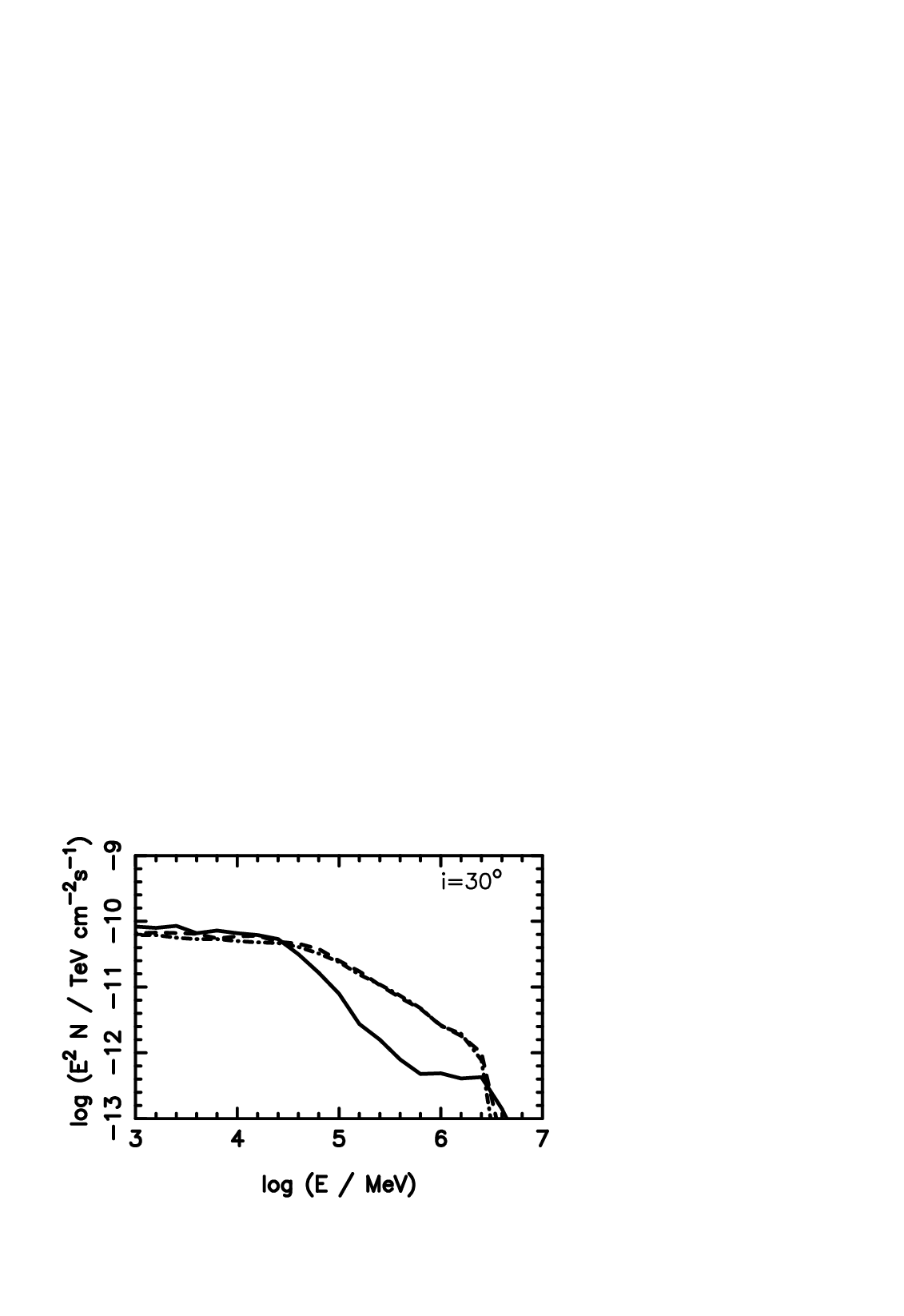}
\includegraphics{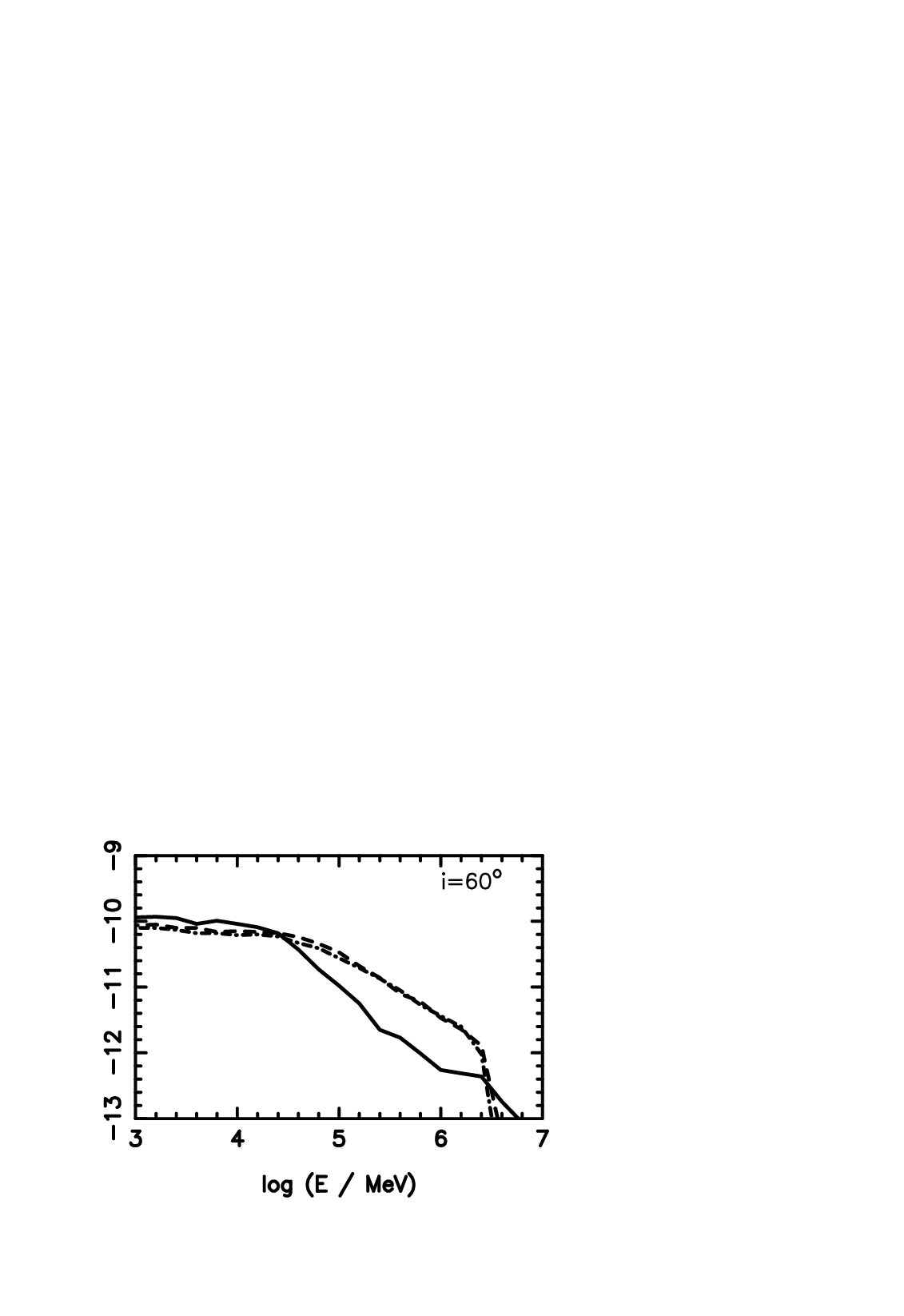}
\caption{The multi-wavelength spectrum observed from LSI +61$^{\rm o}$ 303 
(X-ray data from the ROSAT (R - Goldoni \& Mereghetti 1995) and the OSSE (O - Grove et al.~1995), and in 
$\gamma$-rays from the COMPTEL (C - van Dijk et al.~1996), the EGRET (E - Kniffen et al. 1997), and the detection by the MAGIC telescope (Albert et al.~2006)) are compared on the upper figure 
with the inverse Compton and synchrotron spectra calculated in terms of IC e$^\pm$ pair cascade 
modelling (for $\xi = 0.1$ and $\eta = 0.1$ and the electron spectrum $\propto E^{-2.4}$). 
The IC and synchrotron spectra are shown for the periastron passage (dotted curves), and for 
the phase $0.5$ at which the maximum $\gamma$-ray flux is predicted (dashed curves). 
The $\gamma$-ray light curves (relative units) in the energy range $1-10$ GeV and $>200$ GeV are shown in the middle figures for the inclination angles of the binary system $i = 30^{\rm o}$ (full curve) and $i = 60^{\rm o}$ (dashed). The cascade $\gamma$-ray spectra escaping towards the observer for selected 
phases of the compact object, these two inclination angles and $\xi = 0.1$ and $\eta = 0.1$ 
are shown in the bottom figures: the periastron (full curve) and the apastron passages 
(dot-dashed), and the maximum $\gamma$-ray flux at phase $0.5$ (dashed).}
\label{fig4}
\end{figure}
\section{The case of LS I 61$^{\rm o}$+303}

In order to be consistent with the observational constraints
given by the $\gamma$-ray flux observed from the EGRET source 2EG J0241+6119 
(Kniffen et al.~1997), and recently measured spectrum at TeV energies (Albert et al.~2006),
in our modelling a steeper spectrum for electrons injected into the jet 
has to be applied (photon differential spectral index equal to $-2.4$). We calculate 
the $\gamma$-ray spectra escaping towards the observer for different phases of the compact 
object in order to find out exactly at which phases the highest fluxes of GeV and TeV emission 
should be expected. 
In Fig.~3 the upper panel, the results of cascade calculations 
(for $\xi = 0.1$ and $\eta = 0.1$ and the inclination angle $60^{\rm o}$) are compared with 
the multi-wavelength spectrum of LSI +61$^{\rm o}$ 303. The IC spectra are shown only above 
1 GeV  
due to the technical problems with the calculation of the cascade $\gamma$-rays spectra at 
lower energies (restrictions on computing time). In fact, the spectra should extend also 
through the MeV energy range (observed by the COMPTEL detector) and break at lower energies 
(in the region of hard X-rays) due to the inefficient cooling of electrons in the jet
(as discussed in Sect.~2). The $\gamma$-rays in the GeV-TeV energy range are produced in the IC 
process and
the lower energy photons (from X-rays up to the soft $\gamma$-rays) result from the synchrotron 
emission of electrons accelerated in the jet. We show the spectra at the peak of the TeV 
$\gamma$-ray emission occurring 
at phase $\sim$0.5 (dashed curves) and at the periastron passage (dotted curves) at which 
the lowest TeV fluxes are predicted, after normalisation to the average GeV flux of the EGRET source 2EG J0241+6119 reported in the direction of LSI +61$^{\rm o}$ 303. 
Note that predicted GeV flux could vary by only $\sim50\%$. So such normalization does not
have very strong consequences when estimating the phase dependent TeV $\gamma$-ray fluxes.
The $\gamma$-ray 
spectrum at TeV energies is consistent with the observations by the MAGIC telescope at energies 
above $\sim$200 GeV (the power law spectrum with the differential spectral index
$-2.6$ up to a few TeV, Albert et al.~2006). The MAGIC 
telescope was not able to detect this source at the low emission state (phases close to the 
periastron). According to our calculations this low state should be characterised
by a flatter spectrum and the flux on a level of $\sim$5 lower than at phase $\sim$0.5.    
Note moreover, that the level of synchrotron emission produced at phase 0.5 is similar to the ROSAT X-ray observations
reported by Goldoni \& Mereghetti~(1995) but stronger than the flux reported from direction of 
LSI +61$^{\rm o}$ 303 by the OSSE detector (Grove et al.~1995). 

In Fig.~3 the middle panel, we show the expected
$\gamma$-ray light curves from LSI +61$^{\rm o}$ 303 in the GeV (1-10 GeV) and the TeV energies 
($>200$ GeV). Clear anti-correlation is predicted between the fluxes in these two energy ranges. 
This is due to 
the importance of the cascading processes inside the binary system. 
The lowest TeV fluxes are expected at the periastron passage of the compact object and 
the highest fluxes should be observed after the periastron (with the peak at 
phase $\sim 0.5$). 
This specific phase distribution is due to the absorption effects of the TeV $\gamma$-rays 
(cascading effects) and the location of the observer at the azimuthal angle $\sim 70^{\rm o}$ 
after the periastron. 
Note that the observer located at the  inclination angle $30^{\rm o}$ should observe a 
high level of TeV emission in broader range of phases (half width between 
$\sim 0.4$ up to $\sim 0.1$), than for the inclination of 
$30^{\rm o}$ (between $\sim 0.4$ up to $\sim 0.9$). This feature might serve as a diagnostic 
of the inclination angle of the binary system. The GeV flux reaches a peak at 
phases $\sim$0.15, when the compact object is behind the massive star for the observer located 
at larger inclination to the plane of the binary system (i.e. at $60^{\rm o}$). 
For this location,
the cascade, developing towards the observer, meets the largest optical depths which results 
in the production of large fluxes of the GeV $\gamma$-rays. The GeV flux could vary by 
$\sim$50$\%$ with the 
phase of the binary system. Interestingly, close to these phase ($\sim$0.2) there should also be
observed a small local peak 
in the TeV $\gamma$-ray light curve. This peak is produced by electrons accelerated at larger 
distances from the base of the jet at which the escape conditions of $\gamma$-rays to the 
observer are more 
favorable. Then, TeV $\gamma$-rays are directed at relatively small angles respect to 
the massive star photons which enhances the efficiency of $\gamma$-ray production.

In Fig.~3 the bottom panel, we also show the $\gamma$-ray spectra for a few selected phases of 
the compact object and 
both inclination angles $i = 30^{\rm o}$ and $60^{\rm o}$. It is clear that the GeV fluxes and 
spectra do not vary strongly but the TeV fluxes and spectra ($>200$ GeV) change significantly 
with the 
phase of the compact object. Our calculations indicate that the TeV spectrum at the periastron 
should be a factor of $\sim$5 lower than at the phase $\sim$0.5. 
Moreover, the TeV spectrum at the periastron should be steeper below $\sim$200-300 GeV and 
significantly flatter above this energy respect to the TeV spectrum expected at 
the phase $\sim$0.5.

\section{Discussion and Conclusion}

The massive binary systems provide unique conditions for investigating of the basic problem of 
astrophysics, i.e. the acceleration process of particles, due to the presence of a well-defined 
target (the radiation field of the massive star). However in order to derive useful constraints 
the details of the sometimes complicated radiation processes by accelerated particles 
(e.g. cascades in the anisotropic radiation of the massive star) have to be considered.
Calculations of the type described in this paper predict observational features of the 
$\gamma$-ray emission (in the GeV to TeV energy range) which, if observed, strongly indicate 
on a very efficient acceleration of electrons already close to the base of the microquasar jets 
and allows to constrain the 
efficiency of this acceleration process. In fact, the $\gamma$-ray emission from the microquasar 
type massive binary LS 5039 has been already observed in the TeV energies (Aharonian et al.~2005) 
and in the GeV energies (Paredes et al.~2000), and from LSI +61$^{\rm o}$ 303 is observed 
in the TeV energies (Albert et al.~2006) and  in the 
GeV energies (Kniffen et al.~1997). Our cascade IC $e^\pm$ pair model predicts the $\gamma$-ray 
light curve in the GeV energy range from LSI +61$^{\rm o}$ 303 (Fig.~3) which shows 
maximum close to the periastron as suggested by Massi~(2004). In contrast, the TeV $\gamma$-ray 
light curve peaks at the phase $\sim 0.5$  
showing clear anti-correlation with the GeV emission (see also earlier calculations of the 
propagation of $\gamma$-rays inside LSI +61$^{\rm o}$ 303 by Bednarek~(B06)). 
This cascade model also predicts the flattening of the $\gamma$-ray emission above a few hundred GeV and lower $\gamma$-ray flux at phases close to the periastron.

\begin{figure}
\vskip 4.5truecm
\includegraphics{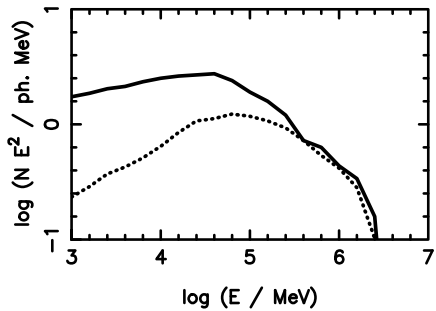}
\caption{The comparison of the $\gamma$-ray spectrum produced by electrons in the
jet directly to the observer (dotted curve) with the $\gamma$-ray spectrum escaping to the 
observer in the cascade process considered in this paper (full curve). The parameters of the example model are the following: $\xi = 0.1$ and $\eta = 0.1$ and the electron spectrum $\propto E^{-2}$, inclination of the binary system $i = 60^{\rm o}$ and the compact object is at the apastron passage.}
\label{fig4}
\end{figure}

The $\gamma$-ray spectra shown in Figs.~2 and 3 are calculated only from the part of the jet 
extending below $10r_\star$ (in order to fulfil the condition of
local, efficient cooling of electrons) and above $0.1r_\star$. 
Note, however, that the considered region of the jet is broad enough as to provide the dominant 
contribution to the escaping $\gamma$-rays (see also Bosch-Ramon et al.~2006). 
The region below $0.1r_\star$ might contribute only to the GeV part of the spectrum. 
Moreover, the IC model in which electrons in the inner 
part of the jet scatter soft radiation coming from the accretion disk should predict different 
$\gamma$-ray light curves. For example, in our model the $\gamma$-ray flux 
depends on the relative orientation of the jet, massive star and the observer predicting 
the peak of TeV emission at phase $\sim$0.5. These features can be 
tested by the future GeV and TeV $\gamma$-ray observations.
As we mentioned in Sect.~1, the IC  model in which collimated electrons scatter 
stellar radiation has been  already considered in a few papers, e.g. recently by Bosch-Ramon
et al.~(2006) , Dermer \& B\"ottcher~(2006). 
These previous works do not take into account the  IC $e^\pm$ pair cascade effects which, 
as we have show above, can play an important role. In order to quantify this effect we compare in Fig.~4 the $\gamma$-ray spectra produced by electrons in the jet directly toward the observer and the $\gamma$-ray spectra which are formed in the cascade process for the apastron passage where 
differences between these two spectra are expected to be the smallest.
Although the $\gamma$-ray spectra above a few hundred GeV are almost this same, they differ significantly at lower energies. Therefore, the broad band shape of the $\gamma$-ray spectra are clearly different. 
Note that similar effects are expected for the inclination angle $i=30^{\rm o}$ since differences between the spectra at the apastron passage for these two inclination angles are small (see the light curves for these two inclination angles in Fig.~3 for the phase of $0.73$).  
The cascade process considered in this paper significantly re-distribute the energy in the 
$\gamma$-ray spectrum toward the observer, influencing e.g.  the  
normalization of the spectrum  at lower energies.
In the works mentioned above the contribution of $\gamma$-ray emission from electrons  scattering of the disk radiation in the jets of microquasars have also been included. It was shown that this contribution can not dominate over the $\gamma$-rays produced  by scattering the stellar radiation. 

On the other hand, electrons accelerated in the jet 
above $10r_\star$ might contribute to the highest energy part of the $\gamma$-ray spectrum and 
to the low energies (X-rays) as a result of synchrotron process and synchrotron self-Compton 
mechanism since synchrotron energy losses of electrons might dominate far away from the massive 
star (e.g. Atoyan \& Aharonian~1999).
However, significant variability of the TeV flux with timescales similar to the period of the binary system has been recently detected by the MAGIC telescope (Albert et al.~2006). 
So then, the contribution to the observed $\gamma$-ray spectrum from the outer parts of 
the jet should be rather minor in case particle injection were constant along the orbit. 
The problem of the phase dependent accretion onto the compact object has not been considered 
in the present work (for possible indications how it can occur see e.g. Bosch-Ramon et al.~2006, Christiansen et al.~2006).
We concentrated only on the radiation processes occurring during uniform injection of energy 
along the jet (independent on the phase). 

The electrons cooled up to a few GeV in the inner jet are advected to the outer parts of the jet. 
They can be responsible for the lower energy emission observed from some microquasars. 
Based on the considered here cascade model, we estimate that a significant 
fraction of the particle energy ($\sim 10^{-3}$) survives the $0.1-10r_\star$ region to allow 
the formation of radio jets, but it's treatment is out of the scope of the paper. 

\section*{Acknowledgments}
I would like to thank Dr G. Romero for useful comments and reading the manuscript
and the anonymous referee for many valuable comments and corrections.
This work is supported by the Polish MNiI grant No. 1P03D01028. 


\end{document}